\documentclass[12pt,unsortedaddress,amsmath,amssymb,preprint,superscriptaddress,
showpacs,preprintnumbers,
]{revtex4-1}%
\usepackage{setspace}
\usepackage{graphicx}
\usepackage{bm}
\usepackage{natbib}
\usepackage{subfigure}
\begin{document}


\title{Non-equilibrium spatial distribution of Rashba spin torque in ferromagnetic metal layer}

\author{N. L. Chung}
\email{g0600187@nus.edu.sg}
\affiliation{Computational Nanoelectronics and Nano-device Laboratory, Electrical and Computer Engineering Department National University of Singapore, 4 Engineering Drive 3, Singapore 117576.}
\author{M. B. A. Jalil}
\affiliation{Computational Nanoelectronics and Nano-device Laboratory, Electrical and Computer Engineering Department National University of Singapore, 4 Engineering Drive 3, Singapore 117576.}
\affiliation{Information Storage Materials Laboratory, Electrical and Computer Engineering Department National University of Singapore, 4 Engineering Drive 3, Singapore 117576.}
\author{S. G. Tan}
\affiliation{Computational Nanoelectronics and Nano-device Laboratory, Electrical and Computer Engineering Department National University of Singapore, 4 Engineering Drive 3, Singapore 117576.}
\affiliation{Data Storage Institute, A*STAR (Agency for Science,
Technology and Research), DSI Building, 5 Engineering Drive 1,
Singapore 117608.}
\date{\today}

\begin{abstract}
We study the spatial distribution of spin torque induced by a strong Rashba spin-orbit coupling (RSOC) in a ferromagnetic (FM) metal layer, using the Keldysh
non-equilibrium Green's function method. In the presence of the $s$-$d$ interaction between the non-equilibrium conduction electrons and the local magnetic moments, the RSOC effect induces a torque on the moments, which we term the Rashba spin torque.

A correlation between the Rashba spin torque and the spatial spin current is presented in this work, clearly mapping the spatial distribution of Rashba spin torque in a nano-sized ferromagnetic device. When local magnetism is turned on, the out-of-plane ($S_z$) Spin Hall effect (SHE) is disrupted, but rather unexpectedly an in-plane ($S_y$) SHE is detected. We also study the effect of Rashba strength ($\alpha_R$) and splitting exchange ($\Delta$) on the non-equilibrium Rashba spin torque averaged over the device. Rashba spin torque allows an efficient transfer of spin momentum such that a typical switching field of 20 mT can be attained with a low current density of less than $10^7$A/$\mathrm{cm}^2$.
\end{abstract}
\maketitle
\section{\label{sec_intro4}Introduction}
Ever since the theoretical prediction of the spin transfer torque
(STT) \cite{slonczewski:jmmm159,berger:9353}, there has been much
research effort in utilizing the STT phenomenon to induce
magnetization switching and precession in ferromagnetic (FM)
nanostructures without the need for an externally applied magnetic
field. Devices which rely on the STT effect for magnetization
switching offer the advantages of lower power consumption and reduced
device dimension, which are crucial factors for nanoscale and
high-density spintronic applications. The STT effect has been
studied in conventional magnetic nanostructures such as spin valves
\cite{katine:prl3149} and magnetic tunneling junctions
\cite{huai:apl84}. For STT to occur in these magnetic multilayers,
one requires a pair of FM layers, i.e., a reference spin layer to
generate a spin-polarized current for injection into the second free
(switchable) layer. The two layers are magnetized in a noncollinear
configuration so as to induce the transfer of the transverse spin
momentum from the reference to the free layer, which is mediated by conduction electrons
flowing between the two layers. In the above process, the role of the spin-orbit
coupling (SOC) effect is neglected. However, it is
well-established that SOC can generate a nonequilibrium spin
accumulation under the passage of current. Thus, it is conceivable
that, in the presence of strong SOC effect, one can induce a STT
without the need for an additional reference FM layer. This is corroborated by
previous theoretical work which showed that the presence of Rashba spin-orbit
coupling (RSOC) whose strength is denoted by $\alpha_R$, and
exchange interaction $\Delta$ between conduction electrons and local spins,
can give rise to domain wall motion via spin momentum transfer
\cite{obata:prb77}. The same spin transfer mechanism can also occur
in a FM layer with a large $\alpha_R$ and $\Delta$ values
\cite{tan:annal326,manchon:prb78}. The predicted RSOC-induced
spin momentum transfer was experimentally demonstrated in a nanowire array
\cite{mihai:natm9,pi:apl97}. The above findings suggest that
by utilizing Rashba-induced STT, one can achieve magnetization switching
within a single FM layer, without an additional non-collinear FM layer.
Such \emph{single layer switching} holds several potential advantages
over conventional STT devices, such as a more symmetric current switching
profile and the reduced influence of spin depolarization at the interfaces.\\

A key element which determines the feasibility of Rashba STT is the presence
of a strong Rashba SOC in the FM metal layer. Initial studies on the Rashba effect
were focused on semiconductor (SC) materials
\cite{miller:prl90,sato:jap89,giglberger:prb75,larionov:prb78,akabori:prb77},
especially in two-dimensional electron gas (2DEG) heterojunction structures, which consist of
two SC layers with different energy bandgaps. The conduction electrons in
the 2DEG experience a strong RSOC effect due to the large potential
gradient, as a result of the band-bending at the heterojunction interface.
However, utilizing the Rashba-induced STT in SC materials is not an attractive proposition as
SCs are intrinsically non-magnetic. Even if ferromagnetic behavior can be induced in them via
doping (e.g. in dilute magnetic semiconductors or DMS), the resulting Curie
temperature lies well below room temperature.
Recent studies have shown, however, that a strong RSOC effect
can also be induced in metallic nanostructures, both of the FM and non-FM types
\cite{lashell:prl77,ast:prl98,cercellier:prb73,krupin:prb71}.
It is known that the Rashba SOC requires a
structural inversion asymmetry (SIA), which gives rise to an internal
electric field. In a metallic FM layer, the SIA can be
enhanced by adjacent layers of heavy metals and oxides, which create
the requisite band structure mismatch and large potential gradient at the interfaces
\cite{premper:prb76,christian:prb77,abdelouahed:prb82,dil.prl101}.
By engineering the interfaces of the metallic FM layer, one can
control the strength of the
RSOC effect within the layer. The ability to enhance the RSOC coupling via
interfacial effects has led to the experimental demonstration of
the effect of Rashba-induced STT, as mentioned previously
\cite{mihai:natm9,pi:apl97}. However, to effectively harness this
effect in future magnetic memory applications, it is essential to
have an understanding of the microscopic spin transport in the presence of the
RSOC effect, and the resulting non-equilibrium spatial distribution of the Rashba-induced STT.\\

Thus, in this paper, we apply the Keldysh nonequilibrium Green's function
(NEGF) technique to study the spin torque generated by the Rashba SOC
on the local magnetization in a metallic FM layer. The NEGF method is
suitable for the study of the Rashba STT, which is essentially driven by nonequilibrium
spin accumulation generated by the passage of current in the
presence of RSOC. In addition, the NEGF method can systematically
incorporate the effects of the leads, and interactions (RSOC and
exchange coupling) as self-energy terms. In Section II of the paper,
we introduce the system Hamiltonian, consisting of the Rashba term
$H_{SO}=\alpha_R(\hat{\boldsymbol{z}}\times\hat{\boldsymbol{p}})$,
where $\hat{\boldsymbol z}$ is a unit vector parallel to the
internal electric field ${\boldsymbol E}$, which acts perpendicular to
the FM layer, $\hat{\boldsymbol{p}}$ is the electron wavevector and
$\alpha_R$ is the RSOC strength. The Hamiltonian also includes the
\textit{s-d} interaction characterized by the exchange energy $\Delta$, which
couples the nonequilibrium spin density due to RSOC effect to the local moments.
Based on the second-quantized form of the Hamiltonian, we apply the
tight-binding NEGF formalism, and calculate various microscopic transport
quantities in the system, such as the local spin current and spin density,
and the overall spin torque generated. In Section III, we numerically
investigate (i) the spin torque efficiency as a function of the strengths
of the RSOC effect $(\alpha_R)$, and the $s$-$d$ exchange interaction ($\Delta$),
(ii) the relationship between the spin torque distribution and the local spin currents,
and (iii) the in-plane spin Hall effect arising from RSOC. Finally,
the summary of results and conclusion are presented in Section IV.

\section{\label{sec_model5}Theory and Model}
  \begin{figure}[ht]
  \centering
  \includegraphics[scale =0.78]{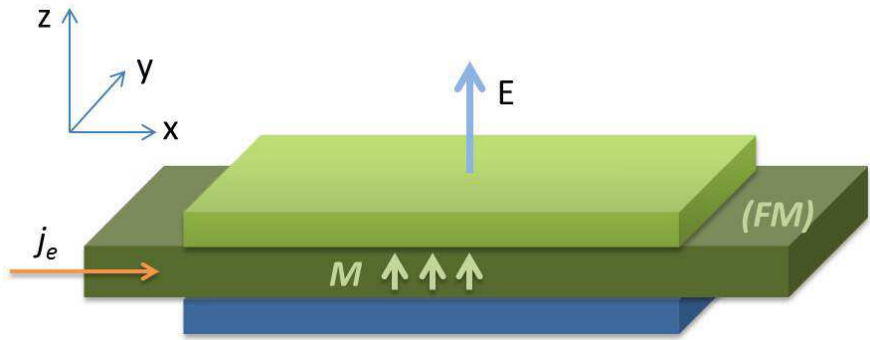}
\caption{Schematic diagram of a ferromagnetic (FM) layer sandwiched
between two dissimilar materials (oxides or heavy elements) to
increase the vertical electric field $E_z$ and thus enhance the
Rashba SOC effect. Current $\boldsymbol j_e$ flows in the in-plane
$x$-direction. The magnetization of the FM layer $\boldsymbol M$ is
oriented in the vertical $z$-direction.}
    \label{fig-1}
  \end{figure}
The structure under consideration is depicted in Fig. \ref{fig-1}.
It consists of a metallic FM layer, sandwiched between two
dissimilar materials (oxides or heavy elements) to enhance the RSOC
interaction at the interfaces and within the FM layer. The local
magnetization $\boldsymbol{M}$ is oriented along the vertical
$z$-direction. A charge current $\boldsymbol{\hat{j}}_e$ is injected
in the $x$-direction, which generates a field $H_{\mathrm{eff}}$
along the
$\boldsymbol{\hat{y}}=(\boldsymbol{\hat{z}}\times\boldsymbol{\hat{j}}_e$)
direction. The Hamiltonian for the system can be expressed as:
\begin{align}
\hat{H}&=\hat{H}_0+\hat{H}_{so},\\
\label{Hkinetic}\hat{H}_0&=\frac{\hat{\boldsymbol{p}}^2}{2m}-\Delta({\boldsymbol{M}}\cdot\boldsymbol{\hat{S}}),\\
\label{HSOC}\hat{H}_{so}&=\frac{\alpha_R}{\hbar}(\hat{\boldsymbol{p}}\times\hat{\boldsymbol{z}})\cdot\hat{\boldsymbol{S}},
\end{align}
where $m$ is the free electron mass, and $\hbar$ is the reduced
Planck's constant. Here, $H_0$ denotes the kinetic energy of the
conduction electrons in the FM layer, ${\boldsymbol{M}}$ is the
magnetization direction, $\Delta$ is the exchange coupling between
the free electron spin and the local moments, and
$\hat{\boldsymbol{S}}=\{\hat S_j\}$ (where $j=\{x,y,z\}$) is the
vector of Pauli spin matrices. $\hat{H}_{so}$ denotes the Rashba
interaction which couples the electron spin with its momentum, with
$\hat{\boldsymbol{p}}$ being the electron momentum, and the
potential gradient inducing the RSOC effect being assumed to be in
the direction $\boldsymbol{z}$ normal to the FM layer. The potential
gradient may arise from a variety of sources such as impurities,
host atoms, and structural confinement
\cite{sih.nat1,kim.apl012504,szunyogh.prl96,castro.prl103,matos.prb81}.
In order to apply the many-body NEGF formalism, the above
Hamiltonian has to be recast into the second quantized form:
\begin{align}
\label{Hkinetic1}\hat{H}_0&=-t_0\sum_{\boldsymbol{r},\sigma}\sum_{\pm}c^{\dag}_{\boldsymbol{r}\sigma}(c_{\boldsymbol{r}\pm\boldsymbol{a}\sigma}+c_{\boldsymbol{r}\pm\boldsymbol{b}\sigma})+\sum_{\boldsymbol{r},\sigma}\varepsilon_{\boldsymbol{r}\sigma}c^{\dag}_{\boldsymbol{r}\sigma}c_{\boldsymbol{r}\sigma}
\\
\hat{H}_{so}&=-it_{SO}\sum_{\boldsymbol{r},\sigma,\sigma'}\sum_{\pm}c^{\dag}_{\boldsymbol{r}\sigma}(\pm
(\hat
S_x)_{\sigma\sigma'}c_{\boldsymbol{r}\pm\boldsymbol{b}\sigma'}\mp
(\hat
S_y)_{\sigma\sigma'}c_{\boldsymbol{r}\pm\boldsymbol{a}\sigma'}).
\end{align}
where $c_{\boldsymbol{r}\sigma}$($c_{\boldsymbol{r}\sigma}^{\dag}$)
is the fermionic annihilation(creation) operator of an electron with
spin $\sigma=\uparrow,\downarrow$ at position $\boldsymbol r$. Here,
$a$ is the lattice spacing representation on a square lattice in the
tight-binding NEGF formulation , $\boldsymbol{a}=a\boldsymbol{e}_x$
and $\boldsymbol{b}=a\boldsymbol{e}_y$ are the unit lattice vectors.
$t_0$ represents the hopping energy between lattice points, and is
obtained by $t_0=\hbar/2ma^2$. The terms
$\varepsilon_{\boldsymbol{r}\uparrow}=4t_0+\Delta/2$ and
$\varepsilon_{\boldsymbol{r}\downarrow}=4t_0-\Delta/2$ represent the
on-site energy at the lattice site, and $t_{SO}=\alpha_R/2a$ is the
SO coupling energy due to the Rashba
interaction.\\

In order to perform numerical analysis through the NEGF, the
retarded ($G^r$) and lesser ($G^<$) Green's functions are required.
These are defined as
\begin{align}
G^r_{\boldsymbol{r}\sigma,\boldsymbol{r}'\sigma'}(t,t')&=i\langle\{c_{\boldsymbol{r}\sigma}(t),c^{\dag}_{\boldsymbol{r'}\sigma'}(t')\}\rangle\theta(t-t'),\\
G^<_{\boldsymbol{r}\sigma,\boldsymbol{r}'\sigma'}(t,t')&=i\langle
c^{\dag}_{\boldsymbol{r'}\sigma'}(t')c_{\boldsymbol{r}\sigma}(t)\rangle
\end{align}
After Fourier transformation, the expression for $G^r$ in energy
space is given by
\begin{align}
G^r(\epsilon)&=[\epsilon-\hat H-\Sigma^r(\epsilon)]^{-1}.\label{Gr}
\end{align}
In the above, $\Sigma^r=\sum_\alpha\Sigma_\alpha^r$ is the retarded
self-energy incurred by the lead $\alpha$, where $\alpha=L(R)$
represents the left (right) lead. $\Sigma_\alpha^r$ can be
determined by $\Sigma_{\alpha}^r=V_{\alpha} g_{\alpha}^r
V^\dag_{\alpha}$, where $V_{\alpha}$ is the coupling matrix between
the lead $\alpha$ and the FM layer, and $g_{\alpha}^r$ is the
retarded Green's function of the lead $\alpha$ and can be calculated
numerically by the renormalization method \cite{nikolic:prb73}.
$G^<$ can be calculated from the relation
\begin{align}
G^<(\epsilon)&=G^r(\epsilon)\Sigma^<(\epsilon)G^a(\epsilon).\label{Glesser}
\end{align}
where $G^a=(G^r)^{\dag}$. $\Sigma^<=\sum_\alpha\Sigma_{\alpha}^<$,
where $\Sigma_{\alpha}^<=if_{\alpha}\Gamma_{\alpha}$ is the lesser
self-energy due to lead $\alpha$, $f_{\alpha}$ is the Fermi function
in lead $\alpha$, and $\Gamma_{\alpha}=-2$Im$\Sigma^r_{\alpha}$ is
the linewidth function representing the coupling between the lead
$\alpha$ and the central FM region.\\

Various transport properties can be evaluated once the different
Green's functions ($G^r$, $G^a$, and $G^<$) have been solved via
Eqs. \eqref{Gr} and \eqref{Glesser}. The charge current through the
system can be expressed in terms of the different Green's functions,
as follows
\begin{align}
I_{\alpha}=&\frac{e}{h}\int^{+\infty}_{-\infty}d\epsilon\medspace
\mathrm{Tr}\left\{\left[\Sigma^<_{\alpha}(\epsilon)A(\epsilon)\right]-\left[\Gamma(\epsilon)_{\alpha}G^<(\epsilon)\right]\right\},
\end{align}
where $A(\epsilon)=i[G^r(\epsilon)-G^a(\epsilon)]$ is spectral
function. Likewise, the current-driven local spin density in the
central region is related to $G^<$ as follows
\begin{align}
\langle s_i\rangle_m&=\frac{\hbar}{2}\sum_{\sigma\sigma'}(\hat S_i)_{\sigma\sigma'}\langle\hat{c}^{\dagger}_{m\sigma}\hat{c}_{m\sigma'}\rangle\nonumber\\
&=\frac{\hbar}{4\pi i}\int^\infty_{-\infty}d\epsilon\sum_{\sigma\sigma'}(\hat
S_i)_{\sigma\sigma'}G^{<}_{mm,\sigma\sigma'}(\epsilon),
\nonumber\\
&=\frac{\hbar}{4\pi i}\int^\infty_{-\infty}d\epsilon\medspace
\mathrm{Tr}[\hat S_iG^{<}_{mm}(\epsilon)], \label{spindensity}
\end{align}
where the subscript $m$ refers to the site index.\\ 

The spin torque exerted on the local magnetization can be defined as the difference between spin current going into and coiming out of the lattice point. We express the spin torque as the divergence of spin current\cite{salahuddin:arxiv}:
\begin{align}
  {\boldsymbol{\tau}}&=\mu_B\int dV \nabla\cdot\boldsymbol j^{S_i},
  \label{spintorque}
\end{align}
where $V$ is the volume, $\mu_B$ is the Bohr magneton, and $\boldsymbol j^{S_i}$ is the spin current density between lattice points. The spin torque ${\boldsymbol \tau}$ can be also be defined as $\boldsymbol{\tau}=-\gamma\boldsymbol{M}\times {\boldsymbol H}$,
where $\gamma$ is the gyromagnetic ratio. We focus on the effective
field $H_{\mathrm{eff}}$ induced by Rashba SOC which acts on the
local moments along the $\boldsymbol{\hat
y}=(\boldsymbol{\hat{z}}\times\boldsymbol{\hat{j}}_e$) direction.
Thus, the effective field due to RSOC is
\begin{align}\label{eq:hcd}
 H_{\mathrm{eff}}&=H_y=\frac{\tau_x}{\gamma M_sV},
\end{align}
where $M_s$ is the saturation magnetization, and $\tau_x$ is
obtained from Eq. (\ref{spintorque}). The torque efficiency is
then given by ratio of $H_{\mathrm{eff}}/{j}_e$.\\

Under steady-state condition and in the absence of dissipative
processes, the spin torque ${\boldsymbol\tau}$, as defined according
to Eq. \eqref{spintorque}, is related to the divergence of
the spin current. By considering the Heisenberg equation of motion,
the local spin bond current between sites $m$ and $m'$ can be
expressed in terms of $G^<$ \cite{nikolic:prb73,hattori:JPSJ78},
i.e.
\begin{align}
\langle\hat{j}^{s_i}_{mm'}\rangle&=\langle\hat{j}^{s_i(kin)}_{mm'}\rangle+\langle\hat{j}^{s_i(SO)}_{mm'}\rangle, \nonumber\\
\langle\hat{j}^{s_{i(kin)}}_{mm'}\rangle&=\frac{et_0}{2\hbar}\int^{\infty}_{-\infty}\frac{d\epsilon}{2\pi}\medspace\mathrm{Tr}[\hat S_i(G^{<}_{m'm}(\epsilon)-G^{<}_{mm'}(\epsilon))]\nonumber\\
\langle\hat{j}^{s_{i(SO)}}_{mm'}\rangle&=[\boldsymbol{e}_{i}\times(\boldsymbol{m'}-\boldsymbol{m})]_{z}\frac{et_{SO}}{2\hbar}\int^{\infty}_{-\infty}\frac{d\epsilon}{2\pi
i}\medspace\mathrm{Tr}[(G^{<}_{m'm}(\epsilon)+G^{<}_{mm'}(\epsilon))],
\label{jbond}
\end{align}
where $(\boldsymbol{m'}-\boldsymbol{m})$ represents the unit vector
between neighbouring sites on the $x$-$y$ plane and $\boldsymbol{e}_{i}$ represents the unit vector of spin $\langle S_i\rangle$. The above
expression for the bond spin current comprises of two terms, i.e.,
the kinetic and SO coupling terms, arising from the corresponding
terms in the Hamiltonian of Eqs. \eqref{Hkinetic} and \eqref{HSOC}.
By considering Eqs. \eqref{spintorque} and \eqref{jbond} together, the spin torque is then given by the divergence of the spin bond current $(\nabla\cdot\boldsymbol j^s)$, which in the discretized tight-binding model is approximated as:\cite{nikolic:prb73,hattori:JPSJ78}
\begin{align}
\label{eq:tx}
\tau_{x(\boldsymbol{m})}&=-\mu_B\left(\langle\hat{j}^{s_z}_{\boldsymbol{m},\boldsymbol{m}+\boldsymbol{e}_x}\rangle+\langle\hat{j}^{s_z}_{\boldsymbol{m}-\boldsymbol{e}_x,\boldsymbol{m}}\rangle\right)/L_{SO},\\
\tau_{y(\boldsymbol{m})}&=-\mu_B\left(\langle\hat{j}^{s_z}_{\boldsymbol{m},\boldsymbol{m}+\boldsymbol{e}_y}\rangle+\langle\hat{j}^{s_z}_{\boldsymbol{m}-\boldsymbol{e}_y,\boldsymbol{m}}\rangle\right)/L_{SO},\\
\tau_{z(\boldsymbol{m})}&=\mu_B\left(\langle\hat{j}^{s_x}_{\boldsymbol{m},\boldsymbol{m}+\boldsymbol{e}_x}\rangle+\langle\hat{j}^{s_x}_{\boldsymbol{m}-\boldsymbol{e}_x,\boldsymbol{m}}\rangle+\langle\hat{j}^{s_y}_{\boldsymbol{m},\boldsymbol{m}+\boldsymbol{e}_y}\rangle+\langle\hat{j}^{s_y}_{\boldsymbol{m}-\boldsymbol{e}_y,\boldsymbol{m}}\rangle\right)/L_{SO},
\end{align}
where $L_{SO}$ is the spin precession length (over which spin precesses by 1 radian), and can be expressed as $L_{SO}=\frac{\pi at_0}{2t_{SO}}$. The above constitutes to the spin torque expression of Eq. \eqref{spintorque}.

\section{\label{sec_result5}Results and Discussion}
Based on the tight-binding NEGF formulation presented in the above
section, we performed numerical calculations of transport parameters
such as the local spin density, bond spin current, and the effective
field $H_{\mathrm{eff}}$ in order to analyze the effect of RSOC
induced non-equilibrium spatial spin torque on the FM layer structure. In our calculations,
the following parameter values are assumed, unless otherwise stated:
$\alpha_R=10^{-11}~\mathrm{eVm}$, $m=9.1\times 10^{-31}$ kg,
$a=0.05~\mathrm{nm}$, $E_F=7.83~\mathrm{eV}$, $M_s=1.09\times 10^6
~\mathrm{Am^{-1}}$, $\Delta=1.6~\mathrm{eV}$ \cite{Miranda.surf117},
and room temperature $T=300$ K. The Fermi energy $E_F$ and
saturation magnetization $M_s$ assume exemplary values corresponding
to that of Co.
\\
\begin{figure}
  \centering
\subfigure{\label{fig:2a}\includegraphics[width=0.5\textwidth]{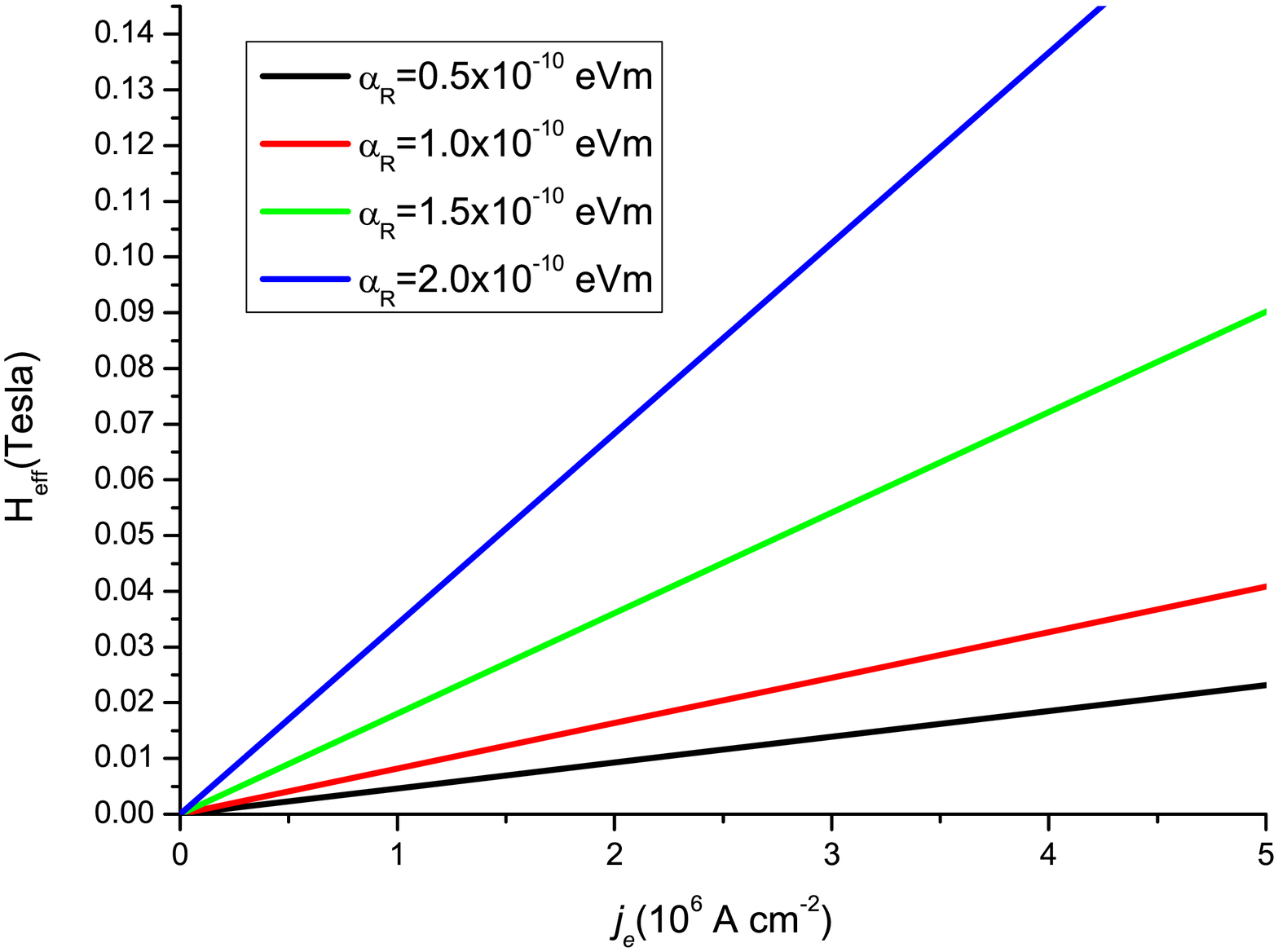}}\\\newpage
  \subfigure{\label{fig:2b}\includegraphics[width=0.5\textwidth]{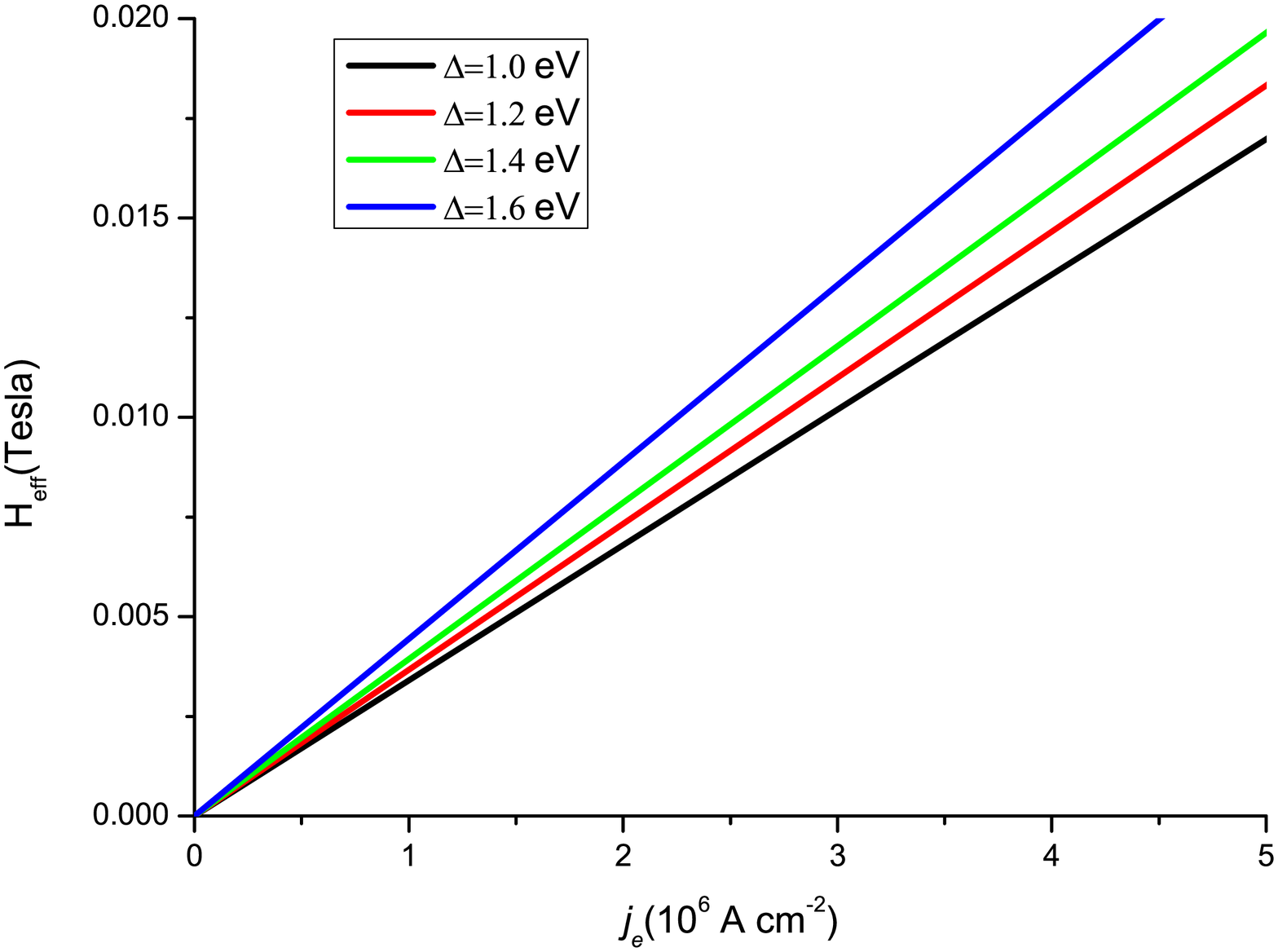}}\pagebreak\\
  \subfigure{\label{fig:2c}\includegraphics[width=0.5\textwidth]{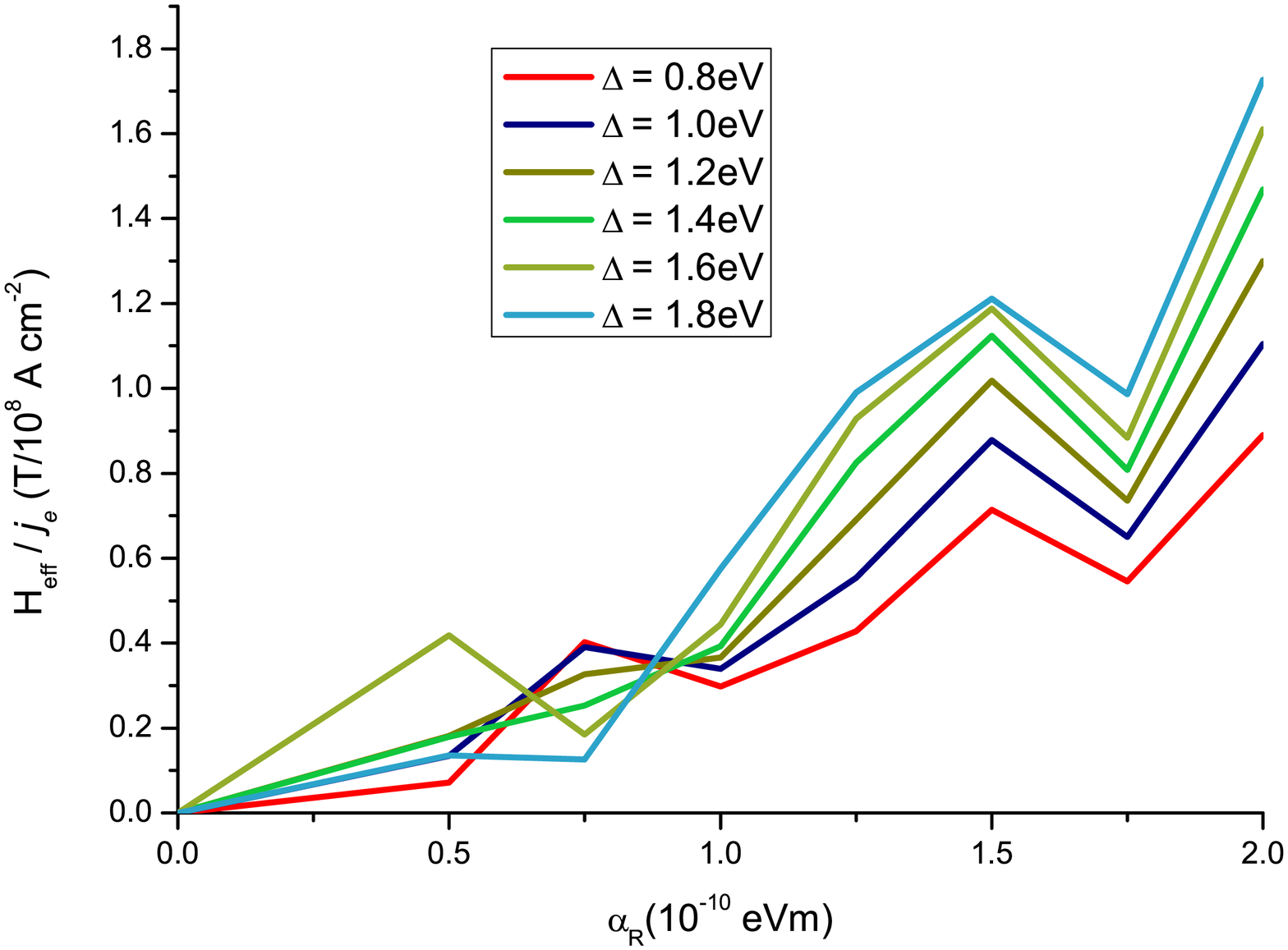}}\pagebreak\\
\caption{The dependence of the effective current induced field
($H_{\mathrm{eff}}$) due to the Rashba spin torque  is plotted as a
function of charge current density (${j}_e$) for (a) varying Rashba
strength $\alpha_R$ with a fixed exchange coupling $\Delta=1.6$ eV,
and (b) varying exchange coupling $\Delta$ with a fixed
$\alpha_R=10^{-10}$ eVm. In (c), the spin torque efficiency
($H_{\mathrm{eff}}/j_e$) is plotted as a function of both $\Delta$
and $\alpha_R$. In the calculations, we assume the dimension of the
sample to be $50a\times 50a$, where $a=0.05$ nm.} \label{fig:2}
\end{figure}
\newpage

We first analyze the role of two key parameters $\alpha_R$ and
$\Delta$ in determining the strength of the effective field
$H_{\mathrm{eff}}$ and the torque efficiency of the system. Figs.
\ref{fig:2a} and \ref{fig:2b} show that, with a fixed $\alpha_R$ and
$\Delta$ respectively, $H_{\mathrm{eff}}$ increases linearly with
$j_e$. This trend is consistent with the prediction that
\begin{align}
 H_{\mathrm{eff}}=\frac{\alpha_R P}{\mu_0\mu_B}(\hat{\boldsymbol
z}\times\boldsymbol j_e),
 \label{torqueformula}
\end{align}
derived from either gauge formulation \cite{tan:annal326} or from
semiclassical (Boltzmann) transport equation \cite{manchon:prb78} in
the strong coupling limit. Eq. \eqref{torqueformula} is a global expression of spin torque under linear response. In the gauge formulation, the factor $P$
assumes a value of $\frac12$ in the adiabatic limit, while in the
Boltzmann model, it refers to the spin polarization of current. We
now consider the torque efficiency, which is given by the gradient
of $H_{\mathrm{eff}}$ with respect to $j_e$. As can be seen from
Figs. \ref{fig:2a} and \ref{fig:2b}, the torque efficiency is
generally enhanced with increase in either $\alpha_R$ and $\Delta$.
However, in our non-equilibrium spatial treatment, it is clear from the plot in Fig. \ref{fig:2c} that the
torque efficiency does not vary linearly with $\alpha_R$, unlike the
prediction of Eq. \eqref{torqueformula}. The difference can be
accounted for by noting that the global expression of Eq. \eqref{torqueformula} is derived in
the limit of large coupling $\Delta$, i.e., up to only the linear
order in $\frac{\alpha_R}\Delta$. In our model, as can be seen from Fig.
\ref{fig:2c}, the torque efficiency shows a slight oscillatory
dependence superimposed upon a general increase with respect to
$\alpha_R$, especially at the region of $\alpha_R<10^{-10}$ eVm. However, at the region where $\alpha_R\geq10^{-10}$ eVm, its behavior is similar to the prediction derived from the Boltzmann semiclassical model for arbitrary coupling strength
\cite{manchon:prb78}. From the effective field $H_{\mathrm{eff}}$,
one can estimate the critical current density required for
magnetization switching. In Figs. \ref{fig:2a} and \ref{fig:2b}, we
consider RSOC strengths ranging from $10^{-11}$ to $10^{-10}$ eVm,
which roughly corresponds to the practical values observed at the
interfaces with heavy metal or oxide layers. Assuming an exemplary
spin polarization of $P=\frac\Delta{E_F}\approx0.5$, RSOC strength
of $\alpha_R=10^{-10}$ eVm, and a switching field of
$H_s\approx0.02$ T applicable for Co nanowire structures
\cite{mihai:natm9}, we find that the critical current density
for switching is approximately $10^6$ A/$\mathrm{cm}^{2}$ [see Fig.
\ref{fig:2a}]. This is significantly lower than the critical current
density of the order of $10^7$ A/$\mathrm{cm}^2$ for the case of the
conventional Slonczewski spin torque in spin valve structures
\cite{jiang:prl92,sukegawa:apl96}.
\begin{figure}
  \centering
\subfigure[]{\label{fig:3a}\includegraphics[width=0.8\textwidth]{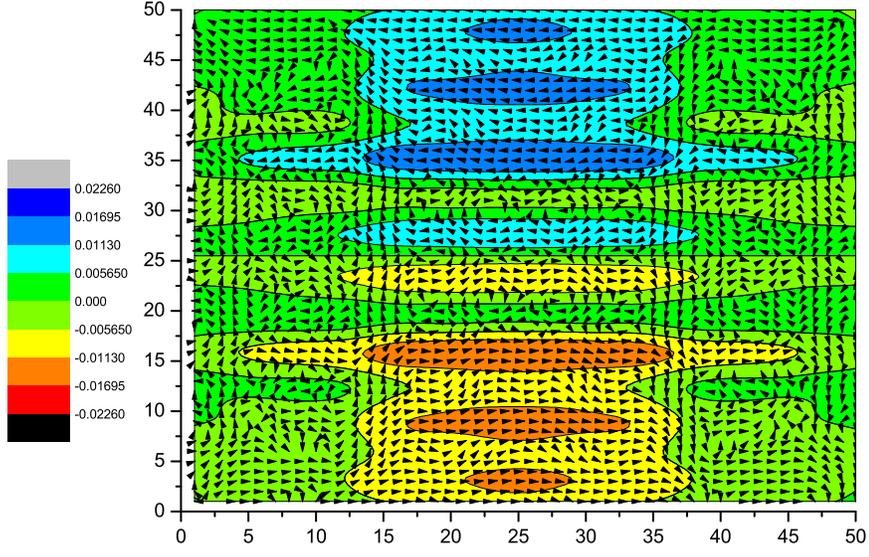}}\\
\subfigure[]{\label{fig:3b}\includegraphics[width=0.8\textwidth]{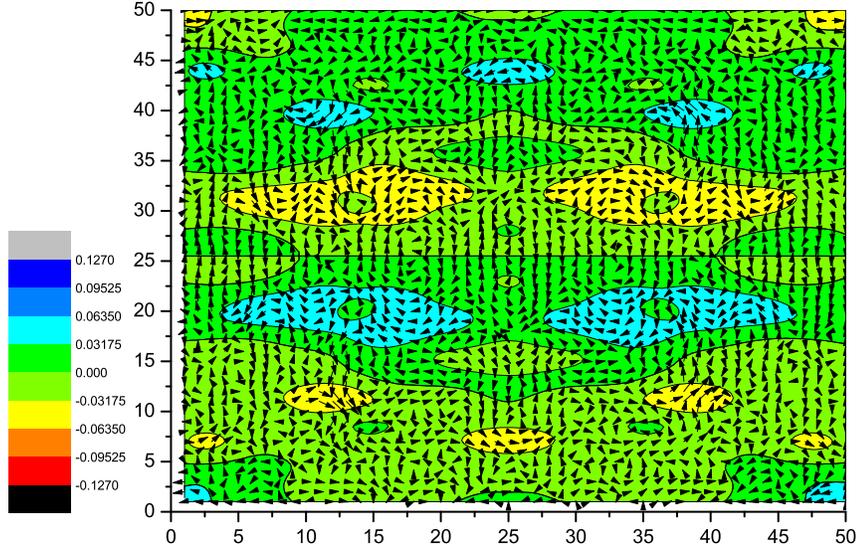}}
\caption{The spatial distribution of the (a) Rashba effect spin
torque $\tau_x$ and its correlation with the local spin current $\langle j^{s_z}_{mm'}\rangle$ by setting $\alpha_R$ to $0.5\times10^{-10}$ eVm, (b) $\tau_x$ and its correlation with $\langle j^{s_z}_{mm'}\rangle$ by setting $\alpha_R$ to $1.5\times10^{-10}$ eVm. The spin torque density is expressed in
units of $\mu_B/L_{SO}$. The sample has a lateral size of
$50a\times50a$.}
\label{fig:3}
\end{figure}
\newpage
Next, we examine the relationship between the Rashba-induced
torque $\tau$ and the spatial distribution of the spin currents. In
Fig. \ref{fig:3a}, we plot the spin torque component $\tau_x$
based on the torque definition of Eq. \eqref{eq:tx}, which
relates it to the divergence of the local spin bond current
$j^{s_z}_{mm'}$. For comparison, we plot the spatial distribution of
the spin bond current $j^{s_z}_{mm'}$ in Fig. \ref{fig:3b}. We observe a
close correlation between the spatial distribution of $\tau_x$ and
the flow of the $z$-polarized spin current $j^{s_z}_{mm'}$. The presence of
RSOC causes a vortex-like flow of the bond spin current
$j^{s_z}_{mm'}$ as shown in Fig. \ref{fig:3b}. Regions where
$j^{s_z}_{mm'}$ is flowing in the $+x$ ($-x$) direction corresponds
to a large positve (negative) $\tau_x$. Conversely, in regions where
the positive and negative spin current fluxes meet and cancel each
other, the spin torque $\tau_x$ becomes small. When the Rashba
coupling strength $\alpha_R$ is increased, the magnitude of $\tau_x$
is generally larger since it scales with $\alpha_R$, as shown in
Fig. \ref{fig:2c}. In addition, the vortices associated with the
spin current become spatially smaller. This may be attributed to the
increase in the rate of spin precession of the conduction electrons
with $\alpha_R$. The increased density of the vortices result in some
cancelation of the bond spin currents near the center of the FM
layer, so that more of the bond spin current flows at the
boundaries, as shown in Fig. \ref{fig:3b}.\\
\begin{figure}
  \centering
  \subfigure[$\Delta=0$, $\alpha_R=10^{-10}$ eVm, $\langle s_z\rangle_m$. SHE recovered.]{\label{fig:4a}\includegraphics[width=0.4\textwidth]{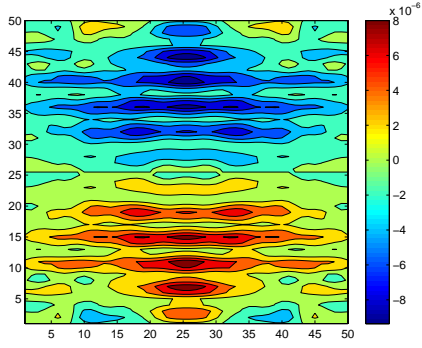}}
  \subfigure[$\Delta=0$, $\alpha_R=10^{-10}$ eVm, $\langle s_y\rangle_m$. Absence of SHE.]{\label{fig:4b}\includegraphics[width=0.4\textwidth]{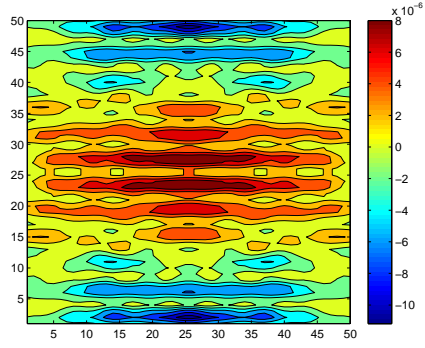}}
 \\
 \subfigure[$\Delta=1.6$ eV, $\alpha_R=5\times10^{-11}$ eVm,  $\langle s_z\rangle_m$. SHE disrupted]{\label{fig:4c}\includegraphics[width=0.4\textwidth]{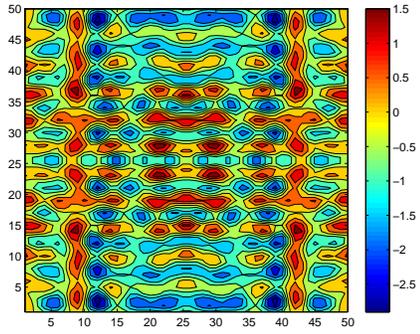}}
\subfigure[$\Delta=1.6$ eV, $\alpha_R=5\times10^{-11}$ eVm, $\langle s_y\rangle_m$. SHE detecteded.]{\label{fig:4d}\includegraphics[width=0.4\textwidth]{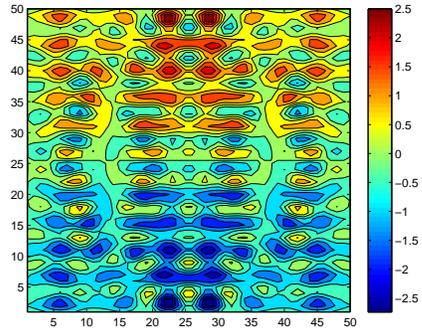}}
\subfigure[$\Delta=1.6$ eV, $\alpha_R=1.5\times10^{-10}$ eVm, $\langle s_y\rangle_m$. SHE disrupted.]{\label{fig:4e}\includegraphics[width=0.4\textwidth]{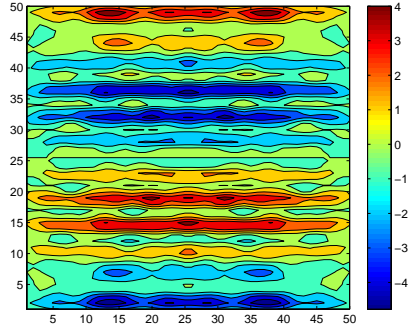}}
\caption{The spatial distribution of the spin density (a) $\langle
s_z\rangle_m$, (b) $\langle s_y\rangle_m$, both with $\Delta=0$ eV, $\alpha_R=1\times10^{-10}$ eVm, (c) $\langle s_z\rangle_m$, (d) $\langle
s_y\rangle_m$, both with $\Delta=1.6$ eV, $\alpha_R=1.5\times10^{-10}$ eVm. In (e) $\langle s_y\rangle_m$ is plotted with a larger $\alpha_R=1.5\times10^{-10}$ eVm, and $\Delta=1.6$ eV.
The sample has a lateral size of $50a\times50a$.} \label{fig:4}
\end{figure}
\newpage
Finally, we analyze the spin density distribution and its dependence
on the exchange strength $\Delta$. Figs. \ref{fig:4a}
and \ref{fig:4b} plot the spin density of $\langle
s_z\rangle_m$ in the absence and presence of $\Delta$, respectively. In the absence of exchange coupling ($\Delta=0$), the distribution profile
of $\langle s_z\rangle_m$ clearly indicates a transverse separation of the $z$-spins,
i.e. an out-of-plane spin Hall effect. This agrees with previous calculations based on the
multimode scattering matrix method which predict a spin-Hall like
separation of the out-of-plane spin component in the presence of Rashba effect
\cite{brusheim:prb74}. However, the clear out-of-plane spin Hall separation
disappears when a sizable exchange $\Delta$ is present, as shown in Fig. \ref{fig:4c}.
It is found that the magnitude of $\langle s_z\rangle_m$ assumes a much larger value throughout the FM
layer. This increase may be attributed to the alignment of the electron spin to the local moments
oriented along the $z$-direction. We also analyze the in-plane spin density
$\langle s_y\rangle_{m}$ distribution, as shown in Figs. \ref{fig:4b}, \ref{fig:4d}, and \ref{fig:4e}. There is no transverse separation of the in-plane
spin density in the absence of $\Delta$ [Fig. \ref{fig:4b}]. This is in line with theoretical prediction where the spin Hall
effect induced by RSOC applies only to out-of-plane spins. However, in the presence of strong exchange
coupling $\Delta$, an ``in-plane" spin Hall effect is present [Fig. \ref{fig:4d}]. This in-plane spin Hall effect
is destroyed in the presence of a strong Rashba strength,
i.e. when $\alpha_R$ is increased to $1.5\times10^{-10}$ eVm [Fig. \ref{fig:4e}]. This may be explained by
noting that a large RSOC strength increases the rate of spin
precession. Thus, the in-plane spin density $\langle
s_y\rangle_{m}$ oscillates and changes signs along the direction of
electron propagation ($x$-direction), as can be seen in Fig.
\ref{fig:4d}. 
\section{\label{sec_conclu5}Conclusion}
In summary, we have studied the non-equilibrium spatial intrinsic spin torque induced by
Rashba spin orbit coupling in a ferromagnetic metal layer. Unlike
the conventional Slonczewski spin torque, the Rashba induced torque
is generated within a single layer, i.e. it does not require spin
injection from an another ferromagnetic reference layer. We analyze
the effect of two crucial parameters determining the strength of the
Rashba spin torque: (i) the strength $\alpha_R$ of the RSOC effect
which is responsible for polarizing the injected charge current, and
(ii) the exchange splitting $\Delta$ which couples the conduction
electron to the local FM moments, thus allowing the transfer of spin
momentum to the latter. The spin transport through the system is
modeled via the tight-binding non-equilibrum Green's function (NEGF)
formalism. The NEGF theory systematically incorporates many-body
effects including interactions with the leads as self-energy terms,
and enables current and spin density to be evaluated spatially under
nonequilibrium (bias-driven) conditions. Based on the NEGF theory,
we numerically evaluate various transport parameters of the system,
such as the effective field $H_{\mathrm{eff}}$ due to the spin
torque, and the spatial distribution of the non-equilibrium spin current and spin accumulation. We found that $H_{\mathrm{eff}}$ generally increases
with both the RSOC strength $\alpha_R$ and the exchange coupling
$\Delta$. However, the dependence of $H_{\mathrm{eff}}$ on both
parameters is not totally linear, unlike previous predictions based
on gauge formulation or semiclassical Boltzmann which are global and only partially non-equilibrium (linear response), and in the strong
coupling limits. For practical values of $\Delta$ and $\alpha_R$,
the calculated critical current density corresponding to a typical
switching field of 200 mT is calculated to be lower than $10^7$
A/$\mathrm{cm}^2$, comparable to that obtained
via the conventional Slonczewski spin torque. For the structure
under consideration where net current is in the $x$-direction and
the local moments are aligned in the vertical $z$-direction, the net
effective field (spin torque) is in the $y$ ($x$)-direction. We plot
the spatial profile of the $x$-component of the spin torque
$\tau_x$, which bears a close correlation to that of the
$z$-polarized bond spin current. It is also observed that the Rashba
torque $\tau_x$ is concentrated near the boundaries of the FM layer.
We also found that the combined presence of RSOC effect and exchange
coupling $\Delta$ induces a Hall separation of in-plane spins,
whereas the spin Hall effect for out-of-plane spins disappear with
the introduction of $\Delta$. Our calculations predict an effective
field $H_{\mathrm{eff}}$ of the order of 1 Tesla for a current
density of $10^7$ $\mathrm{A/cm}^{2}$, thus indicating the
feasibility of utilizing the Rashba induced spin torque to achieve
magnetization switching in spintronic applications.

\end{document}